\documentstyle[fleqn,twoside,gc,cite]{article}

\heads{Vincent Boucher}
      {A cosmological test for general relativity}

\begin{document}
\twocolumn[

\bigskip
\prepno{UCL-IPT-05-01}{\GC {11} 71--74 (2005)}
\Title{A cosmological test for general relativity}

\Author{Vincent Boucher}
       {Institut de Physique Th\'eorique, Universit\'e
	catholique de Louvain, Louvain-la-Neuve, Belgium}

\Abstract
    {The effect of spatial variations of the Newton constant on the cosmic
     microwave background is studied. Constraints on the strong equivalence
     principle violation at the recombination time are then obtained with
     the help of WMAP data and of the standard theory of big-bang
     nucleosynthesis.}

]  

\section{Introduction}

The latest results in cosmography as well as the latest observations of the
cosmic microwave background (CMB) and of supernovae have reinforced the
emergence of a canonical paradigm for cosmology. Most of the cosmological
parameters constituting this concordance model are now known up to five
per cent of relative accuracy. We can rely on these accurate values of the
parameters if the fundamental hypotheses of the canonical paradigm are
checked. Assumptions such as the cosmological principle
\cite{Coles:2003dw,Hansen:2004vq,Hajian:2003qq} and the inflationary
scenario \cite{Coles:2003dw,Bouchet:2004dh} have already been analyzed. But
general relativity is poorly verified at cosmological scales and in the
primordial ages. Deviations from general relativity predicted by the
alternative theories of gravitation are not perceptible today
\cite{Will:2001mx,Williams:2004qb}. However, current experiments dedicated
to cosmology could possibly detect signature of, e.g., the gravitational
sector of string theory. Hence, our interest in testing general relativity
in the early universe is to consolidate the confidence on the cosmological
parameters as well as to probe or constrain new gravitational physics.

Previous works have already studied the impact of alternative theories of
gravitation as pure \cite{Chen:1999qh} or extended
\cite{Catena:2004ba,Nagata:2003qn,Nagata:2002tm} Brans-Dicke theory on the
cosmic microwave background spectrum. Nevertheless, they do not consider
breaking of one of the main features of general relativity: the strong
equivalence principle (SEP).  Only two metric theories of gravitation are
based on this principle, namely, general relativity and Nordstr\"om's scalar
theory. The latter is excluded by the observed light deflection by a
gravitational potential. Consequently, if new gravitational physics exists
beyond general relativity, a strong equivalence principle violation should
be observed (while the Einstein and weak equivalence principles may be
respected).

In this talk I summarize the results obtained in
\cite{Boucher:2004ta,Boucher:2004ub}. The proposed test seeks a SEP
violation in cosmological data for the cosmic microwave background with
the help of the standard theory of big-bang nucleosynthesis. The aim of the
work is to single out and to interpret SEP violation in the CMB power
spectrum due to space variations of the Newtonian coupling.

\section{Strong equivalence principle and its violation}

The strong equivalence principle strengthens the Einstein equivalence
principle in the following way. It extends the universality of free fall for
test particles to compact bodies. Compact bodies are bodies with
non-negligible self-binding gravitational energy. The SEP also enlarges the
independence of non-gravitational experiment outcomes relative to the
location in spacetime where the experiments are performed and relative to
the speed of the freely falling frame in which they are run, to
gravitational experiments. It follows that spacetime variations of the
Newton {\it constant\/} $G$ are sufficient to violate the SEP (but not the
Einstein equivalence principle) since the results of gravitational
experiments would then depend on the position in spacetime.

\subsection{The Nordtvedt effect}

Variations of $G$ induce changing the inertial mass of a given body through
its self-binding gravitational energy. This effect can be implemented, in an
effective way, by adding to the Lagrangian density for compact bodies a
dependence on the new field $G(x)$. It ensues non-conservation of the
energy-momentum tensor
\beq                                          \label{SEPV}
    {\mathcal{L}}_{\rm{m}}={\mathcal{L}}_{\rm{m}}(g_{\mu\nu},\Psi,G(x)
	)\quad \Rightarrow \quad {T^{\mu\nu}}_{|\nu}=\frac{\d
			G}{\d x_\mu}\frac{dT}{dG},
\eeq
where $g_{\mu\nu}$ is the metric and $\Psi$ are non-gravitational fields.
Since in this case the motion of a compact body is no more geodesic, the SEP
is violated. The standard parameterization for space variations of $G$ in
the weak field limit is
\beq                                                  \label{paramG}
	G(\vec x)=G_{\rm N}\left[1+\eta^{\rm gr}\frac{V(\vec
			x)}{c^2}\right],
\eeq
where $V(\vec x)$ represents the external gravitational potential in which
bodies are falling and $G_N$ the value of the Newtonian coupling when $V$
vanishes. The parameter $\eta^{\rm gr}$ characterizes the strong equivalence
principle violation. This parameter equals zero for general relativity and
the Nordstr\"{o}m theory. The non-geodesic motion of massive bodies implies
that the acceleration of these bodies in the potential $V$ depends on their
mass, so that the gravitational mass $m^{\rm gr}$ of a body differs from
its inertial mass $m^{\rm in}$. This is the so-called Nordtvedt effect
\cite{nordt1,nordt2}:
\beq
	m^{\rm gr}=m^{\rm{in}}(1-\eta^{\rm gr}\ s).
\eeq
The difference between $m^{\rm gr}$ and $m^{\rm{in}}$ originates from the
sensitivity $s$ of the mass to variations of $G$; $s$ is also
equal to the compactness of the considered body,
\beq
	s=\frac{\d \ln m^{\rm in}}{\d \ln G}=\frac{|E^{\rm gr}|}
				{m^{\rm in}c^2},
\eeq
$E^{\rm gr}$ being its self-binding gravitational energy.

\subsection{Today's constraints}

Ranging the distance between the Earth and the Moon provides a constraint on
the value of $\eta^{\rm gr}$ today\footnote{Quantities evaluated today or at
the recombination of electrons and protons are indexed with a subscript $_0$
or $_*$, respectively.} \cite{Will:2001mx,Williams:2004qb}
\beq
	\eta^{\rm gr}_0= (4.4 \pm 4.5)\ 10^{-4} \cm {\rm (LLR)}\;.
\eeq

The second kind of test is based on orbital period decay observations of
asymmetric and slightly relativistic binary systems and gives
\cite{Gerard:2001fm,Bailes:2003qc}
\beq
	\eta^{\rm gr}_0 \leq 2.7\ 10^{-4} \cm {\rm (PSR\ J1141-6545)}.
\eeq

The discrepancy between the above constraints and theoretical scenarios
inspired from string theory is solved if the additional gravitational fields
evolve as the Universe expands. In this case, the predicted value of
$\eta^{\rm gr}$, which should be about unity, appears as an initial
condition maintained all over the radiation era. Hence, it is of primary
importance to verify this prediction. In the following sections, we will
look after constraints on the parameter $\eta^{\rm gr}$ at the time of
recombination of electrons and protons contained in the primordial plasma.

\section{Strong equivalence principle in cosmology}

First, let us define our cosmological framework. We assume that the smooth
expansion of a flat universe is governed by the cosmology of Friedmann and
Lema\^{\i}tre. The primordial plasma contains baryons, cold dark matter,
photons and three families of light neutrinos. Moreover, before
recombination, we assume that baryons and photons are tightly coupled by
Compton interactions and thus form a single fluid. Over the homogeneous
Universe picture we add small adiabatic perturbations. The radiation
pressure resistance to compression and rarefaction states of the plasma and
falling of baryons in gravitational potentials generated by dark matter
perturbations, trigger and maintain acoustic oscillations of the plasma.
Temperature fluctuations of the plasma are related to photon density
perturbations. These photons undergo a gravitational redshift at decoupling
of baryons with photons when they climb out of the gravitational potentials.
If we restrict ourselves to first-order perturbations, the plasma behaves
like a set of independent oscillators with different spatial frequencies ---
the Fourier modes. 
This results in a succession of {\it acoustic peaks\/} in the CMB spectrum 
which represents the temperature anisotropy amplitude with respect to the 
inverse angular scale. In gravitational wells (hills), maximum compression 
(rarefaction) states at the decoupling time are associated with odd peaks. 
The weight of baryons shifts the oscillation zero point and appears in the 
CMB spectrum as an increase of the odd peak height. The relative height 
of the first and the second acoustic peaks are then related to the baryon 
density.	

The issue is to compute the spectrum of photon perturbations at the
recombination time in order to put a constraint on $\eta^{\rm gr}_*$. The
oscillation amplitude is proportional to the density of baryons.
More exactly, we should use the gravitational density of the baryon-photon
fluid, since acoustic oscillations originate in the gravitational
interaction between the baryon-photon fluid and dark matter perturbations.
But if the SEP is violated, the gravitational mass of the baryon fluid  
differs from the inertial
one. If $\eta^{\rm gr}>0$, the oscillation amplitude decreases since the
weight of baryons is reduced relative to the inertial baryon density.

To quantify the effect of SEP violation on
the CMB spectrum, we have to evaluate the sensitivity of baryons to 
variations of $G$. For each Fourier mode we consider the falling of an
homogeneous spherical body made of baryons with a radius equal to the
wavelength of the Fourier mode. Therefore the baryon compactness related to
the $n$-th CMB spectrum peak is
\beq
    s^{(n)}_{b,*}\simeq 2.7\ \frac{\Omega_b}{n^2}\ \simeq \frac{0.1}{n^2},
\eeq
where $\Omega_b$ is the ratio of the inertial density of baryons to the
Einstein--de Sitter critical density. Solving \eq (\ref{SEPV}) for the
baryon-photon fluid, we find the following relation between the inertial
density of baryons and their gravitational density $(\Omega_b)^{\rm gr}$ as
measured through the height of the first and second acoustic peaks:
\beq                                                   \label{omegrel}
    (\Omega_b)^{\rm gr}=(1-\eta^{\rm gr}_* s^{(1)}_{b,*})\Omega_b.
\eeq
This determines a cosmological Nordtvedt effect, see \cite{Boucher:2004ub}
for more detail. Determining independently $(\Omega_b)^{\rm gr}$ and
$\Omega_b$ gives a constraint on $\eta^{\rm gr}_*$.

\section{Cosmological constraints}

In this section, we derive two constraints on $\eta^{\rm gr}_*$. Firstly, we
compare the gravitational density obtained by studying the impact of SEP
violation on the acoustic peak height and the inertial density derived from
other observables in the CMB spectrum. Secondly, we assume that the baryon
density provided by CMB experiments is completely dominated by the
gravitational density. The inertial density is then deduced from
measurements of primordial light element abundances.

\subsection{Cosmic microwave background}

The density $(\Omega_b)^{\rm gr}$ is inferred from the height of the
acoustic peaks. The inertial density can be found by analyzing the
horizontal position of the acoustic peaks. Their position depends on the
decoupling time and on the propagation speed of acoustic waves in the
plasma. On the one hand, variation of the gravitational baryonic density due
to the height uncertainty of the acoustic peaks measured by WMAP
\cite{Spergel:2003cb,Bennett:2003bz}, and on the other hand, the permitted
values of the inertial density deduced from the uncertainties about the
horizontal localization of the first acoustic peak allow us to constrain
$\eta^{\rm gr}_*$ \cite{Boucher:2004ta}:
\beq                                       	\label{CMBCMB}
	| \eta^{\rm gr}_*| \leq 0.6\cm  {\rm (CMB-CMB)} .
\eeq
All limits and errors are given at 68\% confidence level. Due to the
approximations we made, this constraint should be taken as to an order of
magnitude.

\subsection{Big-bang nucleosynthesis}

The second constraint on SEP violation comes from a comparison of CMB  and
big-bang nucleosynthesis (BBN) results. Since the density of baryons
extracted from CMB experiments is mainly due to the gravitational density,
we assume the density derived from WMAP, CBI and ACBAR experiments
\cite{Spergel:2003cb,Bennett:2003bz} to be completely dominated by the
gravitational density of baryons, $(\Omega_b)^{\rm gr} h^2=0.022\pm 0.001$.
Production of primordial light element is not directly affected 
by SEP violation since BBN proceeds through non-gravitational interactions. 
Nonetheless, measurements of primordial abundances give direct constraints 
on alternative theories of gravitation (see e.g.,~\cite{Serna:2002fj}). 
Indeed, the production rates of light elements depend on the Hubble rate 
during nucleosynthesis, which is modified by new gravitational physics. 
We choose to compare the WMAP density with the density inferred from the
relative abundance of deuterium and hydrogen $D/H$ which is very sensitive
to variations of the baryon density (see, e.g., \cite{Coc:2003ce}).

From the relative abundance $D/H$ measured through quasar absorption lines,
the baryon content of the Universe equals $\Omega_b h^2 = 0.0214\pm 0.0020$
\cite{Kirkman:2003uv}. Finally, we derive a constraint on $\eta^{\rm gr}_*$
\cite{Boucher:2004ta,Boucher:2004ub}:
\beq
	\eta^{\rm gr}_*\simeq -0.3 \pm 1.0 \cm {\rm CMB-BBN}.
\eeq

\section{Conclusion}

We have proposed a cosmological test for general relativity. This test
probes the strong equivalence principle at the recombination time through
spatial variations of $G$. We have singled out and interpreted the impact of
such a violation on the CMB power spectrum. Two constraints are proposed,
one internal to the CMB, the second using CMB and BBN. No deviation from
general relativity is found. Nevertheless, the constraints do not exclude
the string-inspired value of $\eta^{\rm gr}$ which is about unity during the
radiation era. To derive more confident bounds, we have to perform numerical
simulations that go beyond the simple modification of baryon dynamics
through \eq (\ref{SEPV}) since the gravitational field equations are
modified. We also have to run Monte-Carlo Markov chains to quantify the
alteration of the cosmological parameter values. Hence, we have to work in a
given alternative theory of gravitation with a specific running of
$\eta^{\rm gr}$ and therefore of $G$ in time, thus losing the generic
property of our test.

\Acknow
{This work has been done in collaboration with Y. Wiaux and J.-M. G\'erard. The work of the author was supported by the Belgian Science Policy through the Interuniversity Attraction Pole P5/27.}

\end{document}